# Enhancing immersion in Virtual Reality sports through Physical Interactions

**Arka Majhi**
Interaction Design
M.Des. (2016-18)

Guide: **Prof. Ravi Poovaiah**

**IDC, Indian Institute of Technology, Bombay**

# Approval Sheet

The Interaction Design Project II entitled "Enhancing immersion in Virtual Reality sports through Physical Interactions" by Arka Majhi,
Roll Number - 166330011 is approved, in partial fulfillment of the Master in Design Degree in Interaction Design at IDC School of Design, Indian Institute of Technology, Bombay.

Guide                              :

Chairperson                  :

Internal Examiner       :

External Examiner      :

# Declaration

I declare that this written document represents my ideas in my own words and where others' ideas or words have been included, I have adequately cited and referenced the original sources. I also declare that I have adhered to all principles of academic honesty and integrity and have not misrepresented or fabricated or falsified any idea data/fact/source in my submission. I understand that any violation of the above will be cause for disciplinary action by the Institute and can also evoke penal action from the sources which have thus not been properly cited or from whom proper permission has not been taken when needed.

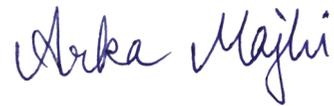

**Arka Majhi**

166330011

IDC, Indian Institute of Technology, Bombay

November 2017

# Acknowledgement

I would like to thank Prof. Ravi Poovaiah for guiding my project. I would like to thank all the professors of interaction design Prof. Anirudha Joshi, Prof. Venkatesh R, Prof. Girish Dalvi and Prof. Jayesh Pillai and all the other professors of IDC whose courses have helped me greatly in doing this project.

I would like to thank the countless Youtubers and Online bloggers for helping me understand the concepts and technologies and make the working prototype for this project.

Finally I would like to thank all my interaction design batch mates and students from M.Des and PhD course for their participation in user studies, usability evaluation, feedback, emotional support and the motivations.

A lot of thanks for the workshop staffs of IDC for their continuous support in working hand in hand and fabricating the prototypes.

Arka Majhi

November 2017

# Abstract

Recent discoveries in VR have opened up scope for designing physical tools and controllers to enhance immersion, through perceived reality. In a virtually simulated sports scenario it is challenging to immerse user because most of the available controllers are unable to bridge the user experience in the real world to the actions in the virtual world.

My research is to identify HCI problems in existing VR controllers, design a physical controller prototype with 'realistic tangible mapping', trying to solve the existing problems and evaluate it in a designed VR game for skating. Its immersiveness would be graded on Likert's scale on parameters like perceived interactivity and reality, spatial presence and enjoyment. The evaluation will be done after trial runs and feedback sessions by playing the game with the designed controller and comparing it with ones available in the market.

The findings will help people understand what all parameters we should consider while designing futuristic controllers, customized for a particular sport.

# Table of contents

# 1. Introduction

Being a social creature, humans find physical/ tangible interaction and human-to-human presence essential for life [1]. VR sports can also supplement that, by allowing virtual controls and imaginative gameplay. However, present VR based sports is often limited by the lack of tangible interaction. Although considering the advancement of controllers, they still are considered as external devices for players and they cannot immerse themselves in an intuitive way. Modern researches suggest that more is the level of body movement sensed and simulated, by a game controller, the more strong and effective is the engagement levels of a player [2,3].

I wanted to extend the interactivity and hence the experiences of mobile VR by introducing additional hardware like controllers with embedded sensors and actuators for sports scenarios.

Since, the evolution of smartphone, with every release it improved its sensors and actuators. It could be foreseen that something similar is going to happen with mobile HMDs that will extend their sensing and actuating capabilities. As smartphones, have no physical interactive sensors, they fail in creating or sensitizing spatial presence [4].

A lot of academic researchers are working on controller's naturalness and VR games. The traditional game theories, help explaining the processes behind enjoyment of games, played with the cutting-edge VR gaming technologies.

## 2. Design Brief

The research aim of my design project is to find and analyze the specific interaction challenges which could improve the experience of sports in VR and try to address those by proposing feasible interaction models through working research proto- types. My study builds upon previous contributions in designing motion control devices from the past.

Through user trials and feedbacks, the experience of sports in VR, played by the new research prototype controller is evaluated against the controllers/joystick already available in the market. Evaluation will be done comparing perceptions of interactivity and reality, presence in space, and delightness in Likert Scale (Ordinal data type). My design project scopes on interactions for end-users/ consumers using VR for Sports applications.

## 3. Background and literature review

Playing games in VR has always been more entertaining and delightful activity than playing in static 2D screens. VR is successful as an immersive gaming platform because it creates alternate realities of the environment for players to explore [5,6]. The immersiveness experienced by players is directly related to the 'play theory', which provides the guidelines for designing, enjoyment for VR games.

Playing games in VR has always been more entertaining and delightful activity than playing in static 2D screens. VR is successful as an immersive gaming platform because it creates alternate realities of the environment for players to explore [5,6]. The immersiveness experienced by players is directly related to the 'play theory', which provides the guidelines for designing, enjoyment for VR games. In VR gameplay, the player has a power of controlling and influencing the gameplay and the virtual environment [7,8]. Through interactivity, the player heads towards achieving an alternate reality. This sensation becomes so strong within, that the player feels effective through his gestures in real life. Vorderer's theory of play states that, the more interactively a player plays in virtual world, the more the player gets deeply immersed in the game world because the player develops an urge to continue [6]. This phenomena leads to a greater sense of presence. Presence (spatial presence to be precise), is the sense of being located inside a abstracted simulated world [9]. The game world envelopes the player's sense of self to an extent that, it overtakes the consciousness of real world .

Perceived realism are categorized into internal and external by Busselle and Bilandzic. In internal realism, the virtual world does not violate the laws of the real world, while in external realism it deviates to certain extent; External realism is introduced to help user easily detect, the unrealistic elements without hampering the experience of realism. VR sports experiences should be so real, that the interaction of players is more viable. Hence, players can deeply immerse into the virtual world, strengthening a sense of participation [10].

Kinect controller interface is facilitating hands free gameplay but still considered intangible because the player cannot physical connect himself with the virtual environment.

Tangibility and materiality of the interface, whole body interaction, embedding of the interface, users' interaction in real spaces and contexts bring in more immersive experiences [11]. Interaction in the virtual environment by physical hand gestures like grasping and moving real world objects makes experiences believable [12,13]. Researchers found that tangible interfaces produce more intuitive interaction [14]. The real body motions simulated into a virtual environment games can be perceived by one as oneself [15].

A few researches have found that more is the interactivity, more it leads to immersiveness. McGloin experimented interactivity in the form of controller type as the present investigation expects to do [16,17]. They found that all the more normally mapped controllers prompted for a more prominent feeling of presence [18]. Persky and Blaskovich exhibited that more interactive is the framework (an Immersive Virtual Environment Technology System, IVET) the higher is immersiveness [19].

Shafer and colleagues showed strong relationship between interactivity and spatial presence [20]. It is safe to assume, at this point, that perceived interactivity leads to immersiveness in virtual world.

# 4. Primary Research

As a user, I've been fascinated by virtual reality's user experience since the first time I tried it. The experiences were so engaging that I felt like living in those virtual experiences. My curiosity about this effect is the reason why I go to every VR demo that I can find. The last time I attended Vamrr-Mumbai, I started identifying some patterns in the way users learned to use the controllers while I was waiting in line. One of those patterns had to do with the user's previous gaming experience: experienced gamers would learn faster how to play.
I compared Oculus Rift and HTC Vive in public demos and in my department, IDC, IIT Bombay. These are some pain points which I found.

## 4.1. Oculus (Touch) pain points

- People who did not have gaming knowledge before attempting Oculus experienced difficulty utilizing the controllers and getting used to the new visual experience. They exhausted additional time learning and less time really playing

- Controller was designed with the human hand in mind and holding them gives a deeply satisfying experience. Fingers naturally fall in to place over the buttons and joy sticks that they correspond to

- It was exceptionally hard to distinguish which buttons to press as shown in the screen

- The initialization process that suggested indicating with a finger to pick a category was the most befuddling. Most of the users failed to recognize which button to press initially, and where to point next

- Although majority of the games had an onboarding instructional exercise, most of the players still required out side guidelines from the demo facilitator before beginning to play

## 4.2. HTC Vive (Wand) pain points

- People figured out how to use the HTC Vive controllers more rapidly than with Oculus Rift. They have just two

buttons —one in the back that is generally a trigger, and one for the thumb for scrolling. Sometimes, the left and right hands do distinctive things. For a few players, perceiving which one does what was troublesome

- Though the urge to learn and adapt about HTC Vive was less steep than Oculus Rift, players still required outside direction to figure out how to use the controllers

- Selecting objects was the most confounded task for new players

- Sometimes players need to press a button in VR with body gestures. In a situation, when we need turning on a light, one need to do it by pointing arm straight to the switch. None anticipated that it would need their arms; they imagined it could be done only with controller

- All clients required outside direction to figure out how to choose certain things in the VR experience; though once they comprehended it, it was effortlessly reused

- Players with previous involvement in VR gaming required less instructions —just a short instructional exercise were enough for them

## 4.3. Lessons learnt

It is tough to conclude which among this two controller is better, but the Vive wands give a slight reminder that one is holding something in hands, while Oculus Touch is successful in melting away as one engages with a given VR experience. In my perception, research and experience in VR, I've prepared some recommendations for enhancing the VR experience using these controllers:

- Minimize controller buttons: Design as few buttons as could reasonably be expected, and ensure each
button is special so that the player does not get confounded

- Unify actions done with a controller: Map the thumb's interaction to scroll and the index finger's interaction to press. By changing these, it would be simpler to distinguish which interaction is requesting by the frequently

# 5. Secondary Research

VR is already cool but to make it truly immersive, one not only need to have their head in another world, but the hands too. Oculus Touch is one of those controllers which make it possible. The designing process of Oculus Touch is well documented and a good reference to see before designing a game controller.

## 5.1. Design process of Oculus touch

Oculus Touch make 'hand presence' possible in VR. Designing it was a process of more than two years of smart thinking and scores of prototypes.

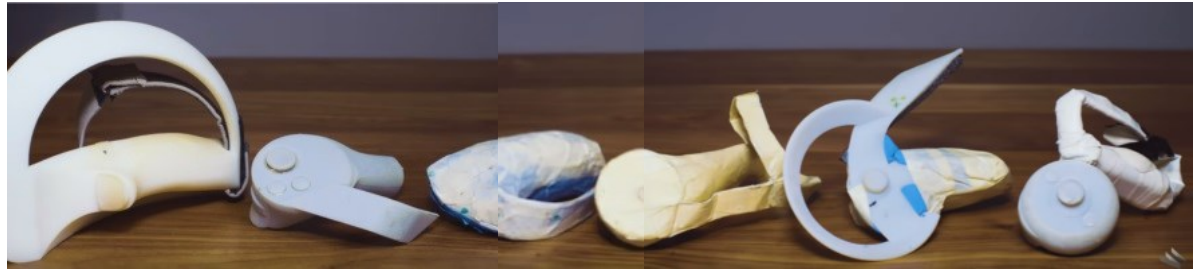

Source : wired.com/2016/12/oculus-touch-design/

In order to bring your hands into VR, there is a need of a perfectly tracked solution, much like a headset which has embedded infrared LEDs, that a sensor reads, tracks and positions something in space.

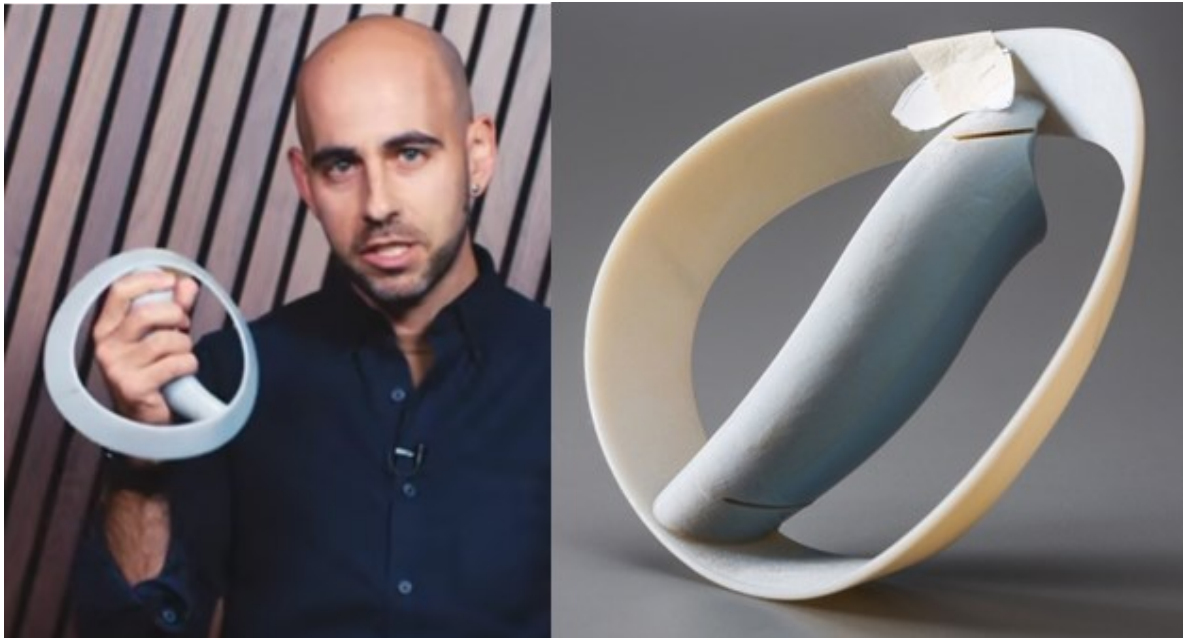

Source : wired.com/2016/12/oculus-touch-design/

This is what it would look like if we made it as easy as possible for the computer. But no one wants to hold a pair of school bus size steering wheels in their hands in order to do something.

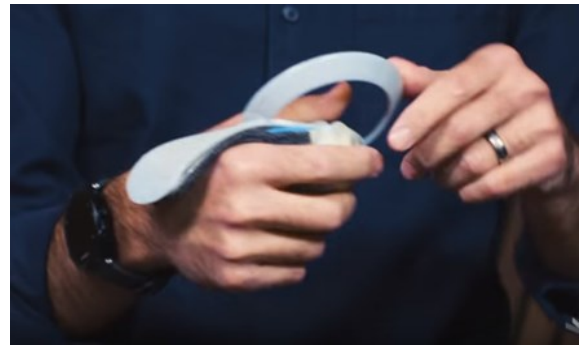

Source : wired.com/2016/12/oculus-touch-design/

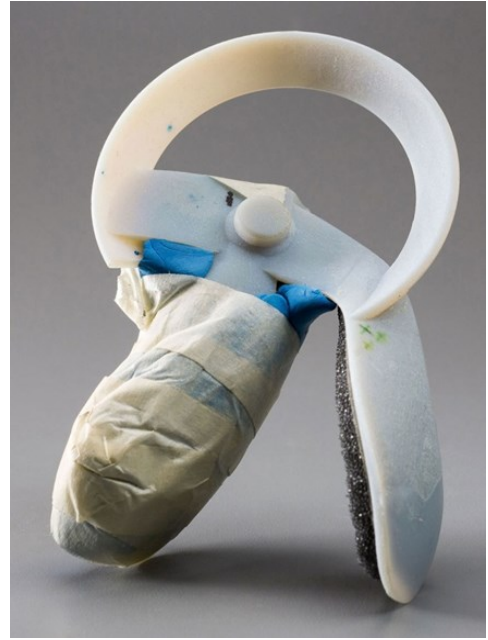

Source : wired.com/2016/12/oculus-touch-

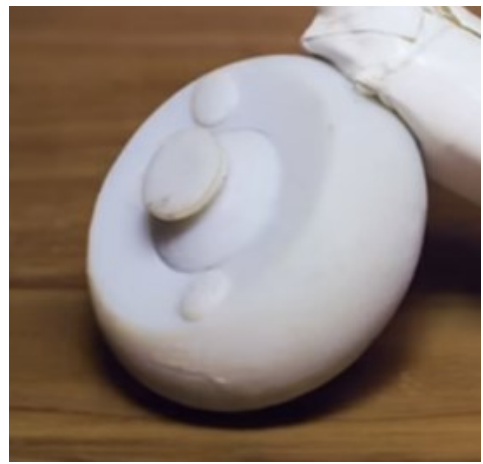

Source : wired.com/2016/12/oculus-touch-design/

As the tracking portion shrunk, the question of how it fit in your hand became something to explore. Was it something one can fit onto own hand or was it something that one could hold .

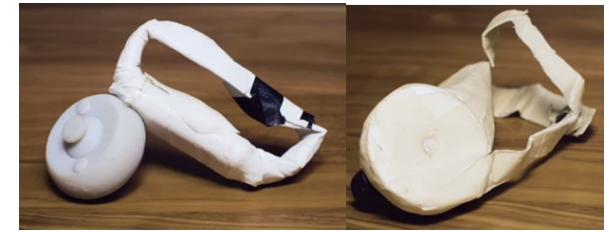

Source : wired.com/2016/12/oculus-touch-design/

The decision was to be taken if something one is going to hold or wear. When one wears something, it has the benefit of being able to just open your hand and it doesn't fall down. But getting these on and off would become a bad user experience

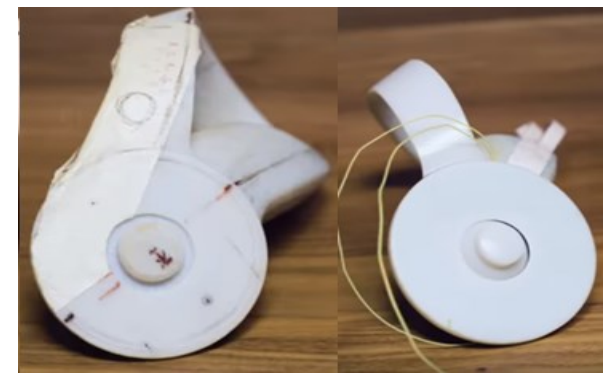

Source : wired.com/2016/12/oculus-touch-design/

While that debate was being held, the decision of what to put on the dial/face was going on. Was it going to be thumb sticks or regular game controller buttons.

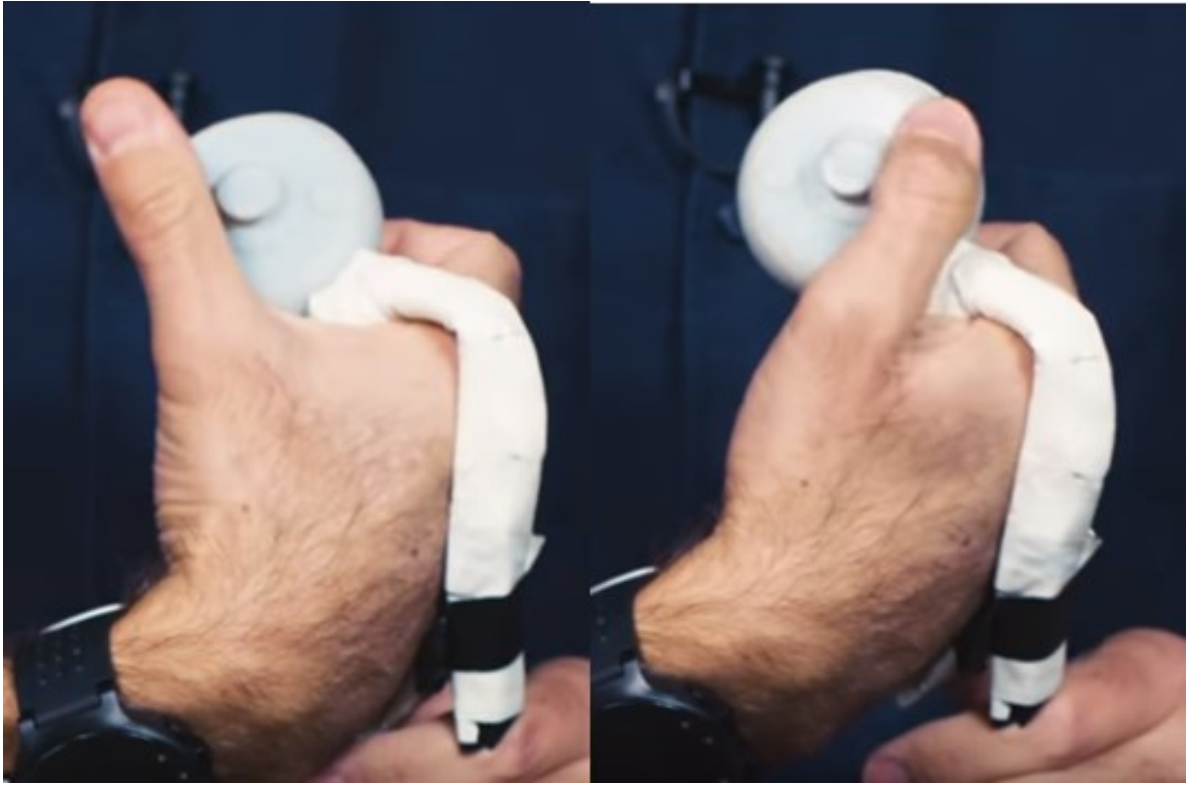

Source : wired.com/2016/12/oculus-touch-design/

This was the first face prototype of the controller. It has an analog thumb stick much like one in a conventional game controller. The Oculus design team did some design thinking and realized that, it wasn't going to work. Our thumb swings out a lot more easily than it swings in. So if there is a centered thumb stick and buttons on either side, that would not be comfortable over long stretches of time.

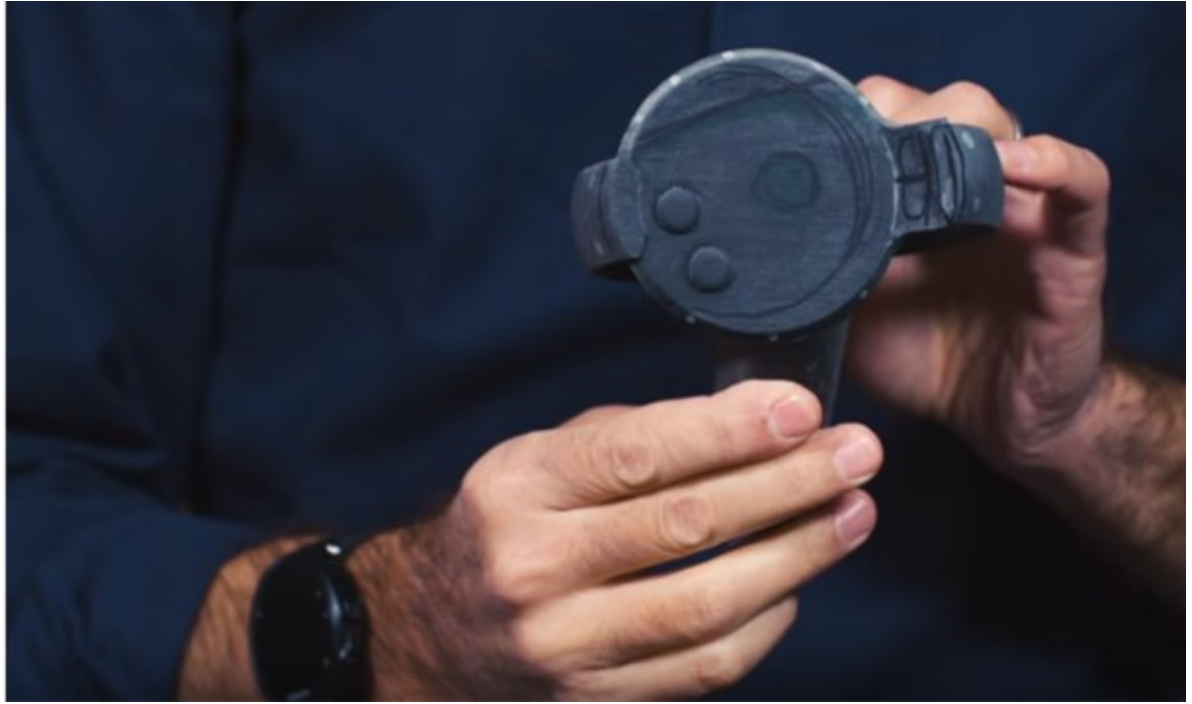

Source : wired.com/2016/12/oculus-touch-design/

These exploration led to the culmination of a lot of decisions and once they got to this prototype, the real design task begun.

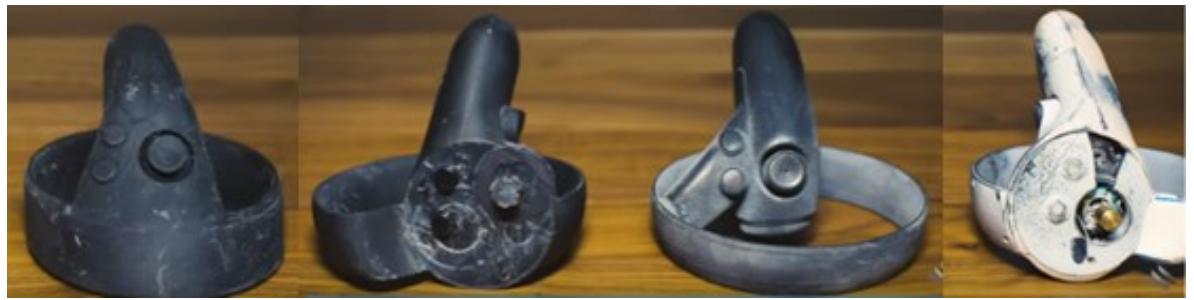

Source : wired.com/2016/12/oculus-touch-design/

The improved design has a small ring with all the embedded tracking on it. It has a two triggers on the body but on the face, the thumb stick is more to the inside and the buttons are to the outside.

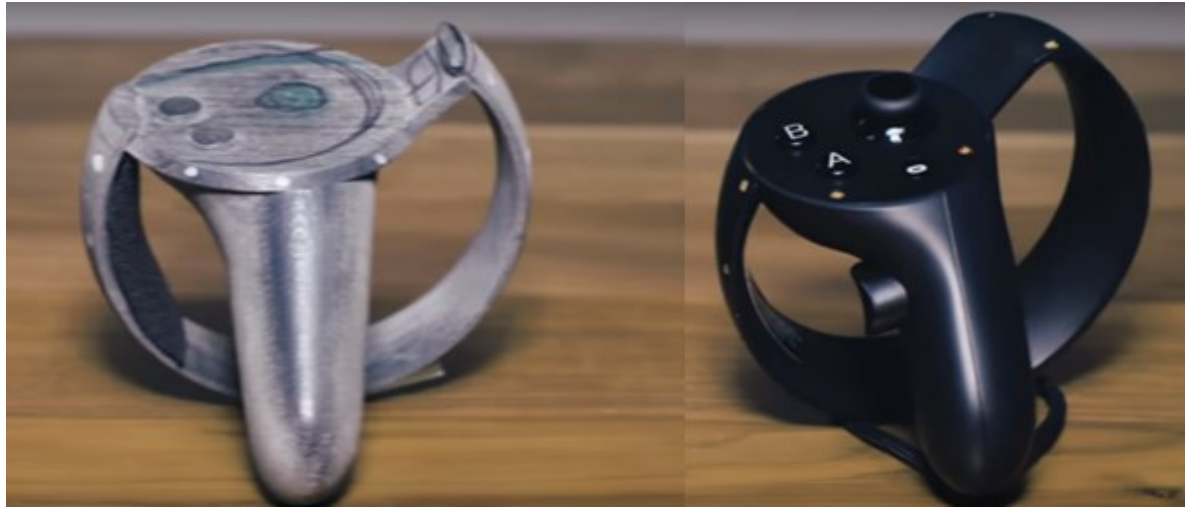

Source : wired.com/2016/12/oculus-touch-design/

The tracking is exposed so that one can see where those LEDs are. The buttons, thumb rest and the thumb sticks all senses when thumb is on them. The triggers register when one presses them. So through various permutations of placing thumb on something resting away from it actuates the hands in virtuality like giving thumbs up, rip things and that's the kind of thing that actually gives 'hand presence'. It's not only about opening or closing of hands, it's about giving our hands a set of discrete actions that helps us, translate gestures you're trying to communicate into a VR surrounding. This story was about two and a half years of design thinking of making the Oculus Touch.

# 6. Designing skateboard controller

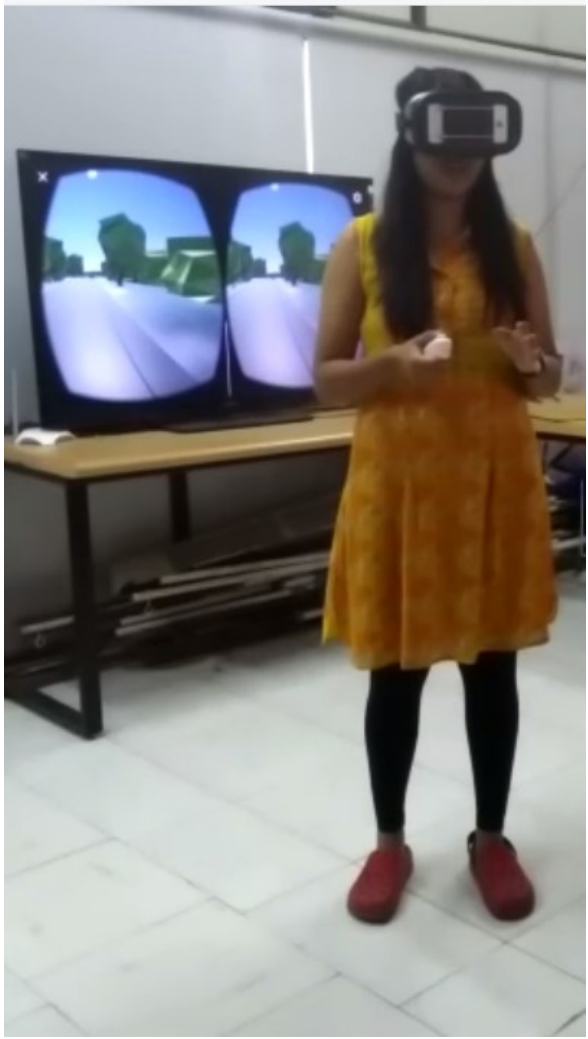

Source : Author

Throughout the development of the controller, I explored how to integrate technology into the human experience of the sport: Skateboarding.

## 6.1. Designing from User Testing

It started with a PC game and a basic VR game with general FPS controls with keyboard and mouse inputs. Students and children were tested with this method. Though they were quite interested with regular controls, but they felt like they are walking or running, rather than skating. Skateboarding is all about performing different stunts as a show off of player's skills. It could be best experienced by standing on a real skateboard and performing those stunts.

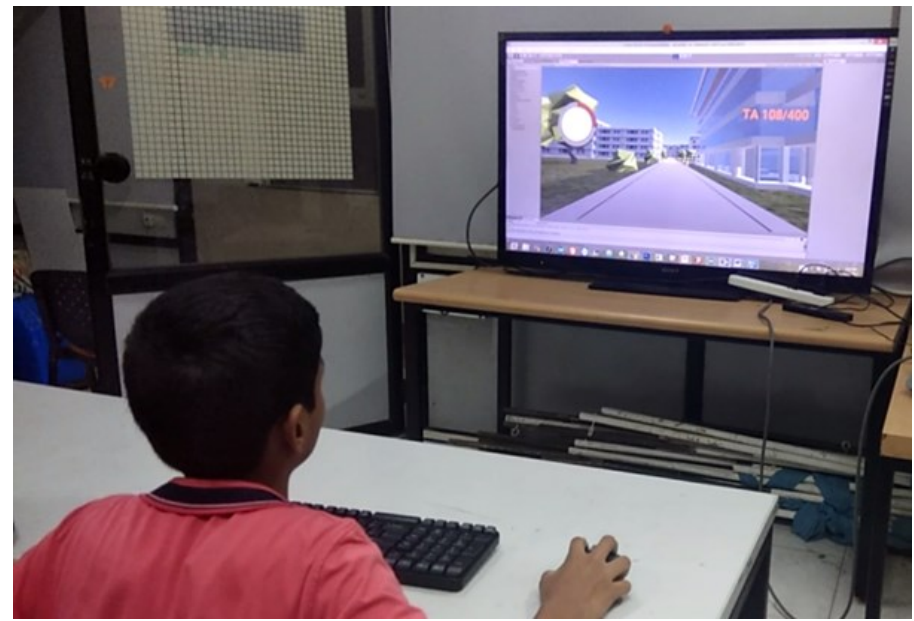

Source : Author

I thought of making a skateboard sense, player's different actions/gestures and perform them in the virtual world

Having previous experience in hardware prototyping with Arduino microcontroller, I thought making Arduino work as HID (Human Interface Device), like an actual PC keyboard, we can control the gameplay character in VR. It would work the same way as playing with a PC keyboard.

But in this case, for simplicity of understanding; keyboard arrow keys are mapped with tilt sensor and spacebar is mapped with jump sensor. This made the controller setup mimic the experience of actual skateboard in virtual world.

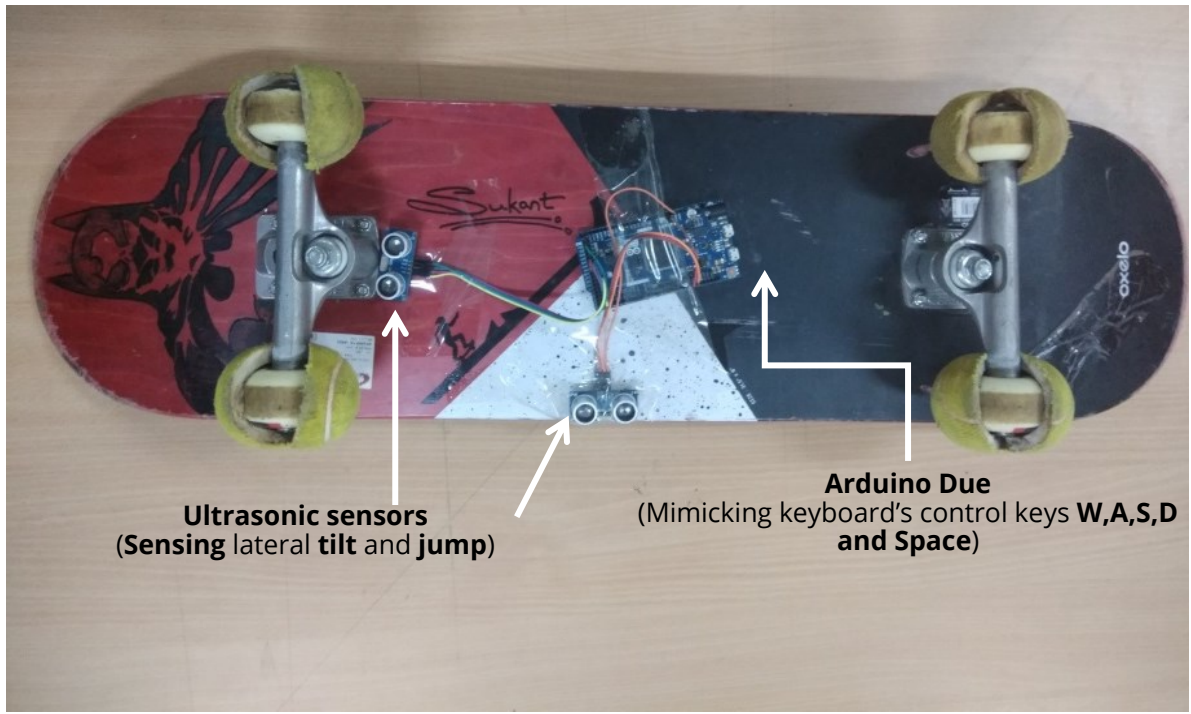

**Ultrasonic sensors**
(**Sensing** lateral **tilt** and **jump**)

**Arduino Due**
(Mimicking keyboard's control keys **W,A,S,D and Space**)

Source : Author

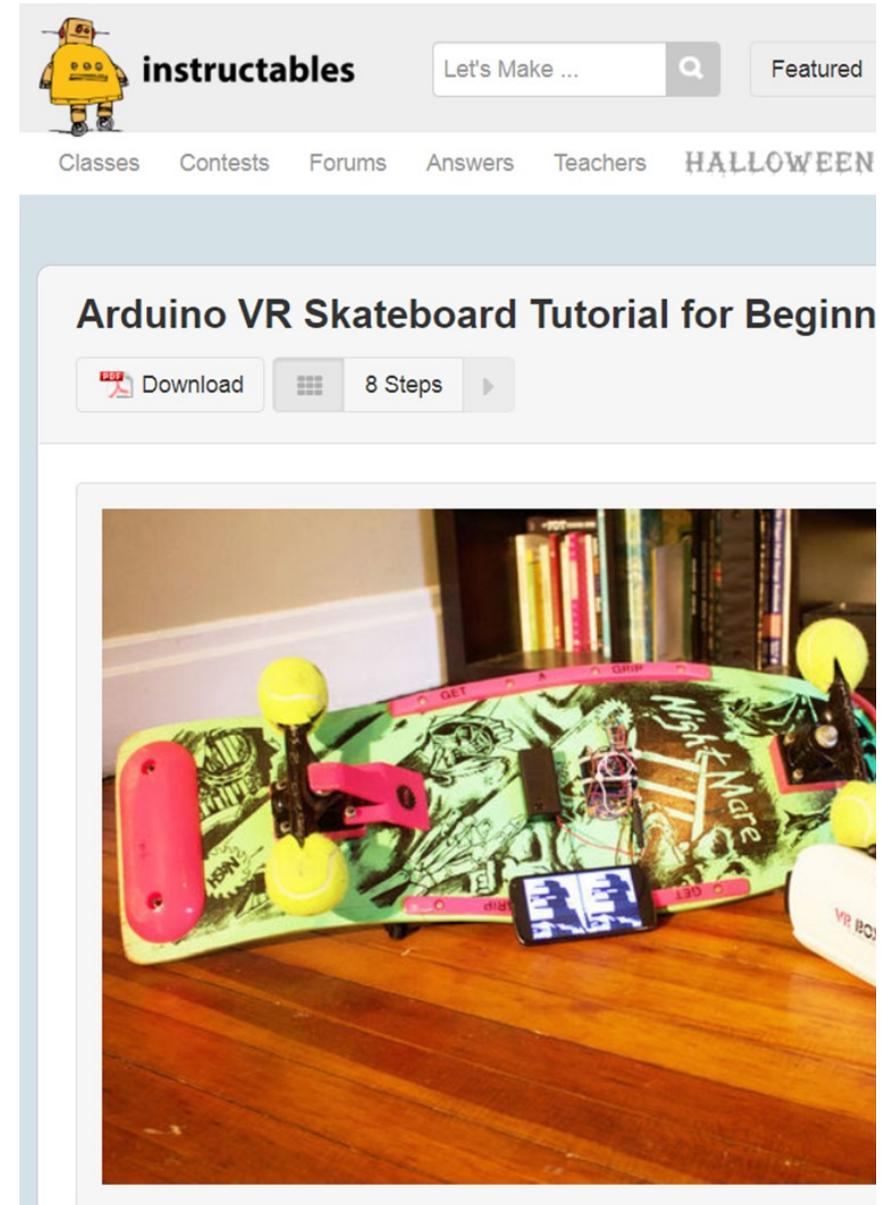

Source : instructables.com

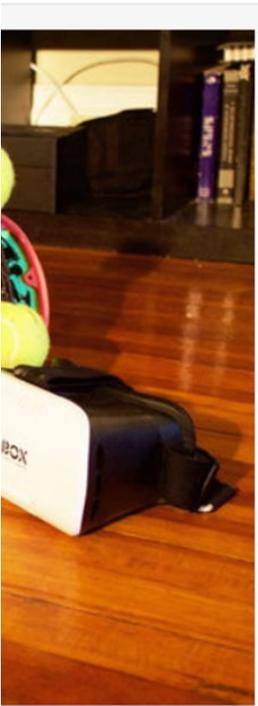

The skateboard controller is embedded with ultrasonic sensors which measured the distance between the skateboard and the ground plane. They are placed one at the longer side and another near to the front wheels. The side sensors check the horizontal tilting. As the player tilts left or right the distance measured by ultrasonic sensor changes. This data is then sent, to the Arduino microprocessor which checks if the value crossing the lower or higher thresholds, previously set in its program code. According to it, it sends instructions to the HMD to move straight, lean left or right. Same thresholds are set for jump sensor, which detect if the skateboard is tilted upwards and triggers for a jump action.

These sensors are very cheap and easily available, but are still considered somewhat inaccurate, compared to their expensive and bulky competitors. However their accuracy is sufficient for general consumer applications.

In VR sports, the player requires to change angular orientation also. It

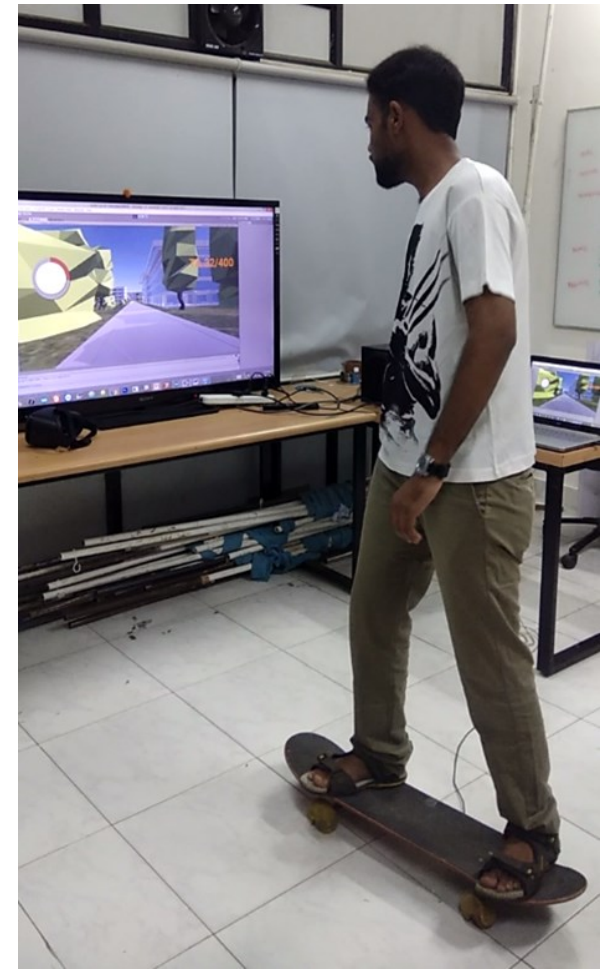

Source : Author

was not possible in only skateboard controller as its rollers are jammed to avoid slipping while eyes are covered in HMD. So it required a turntable with railings to grip on.

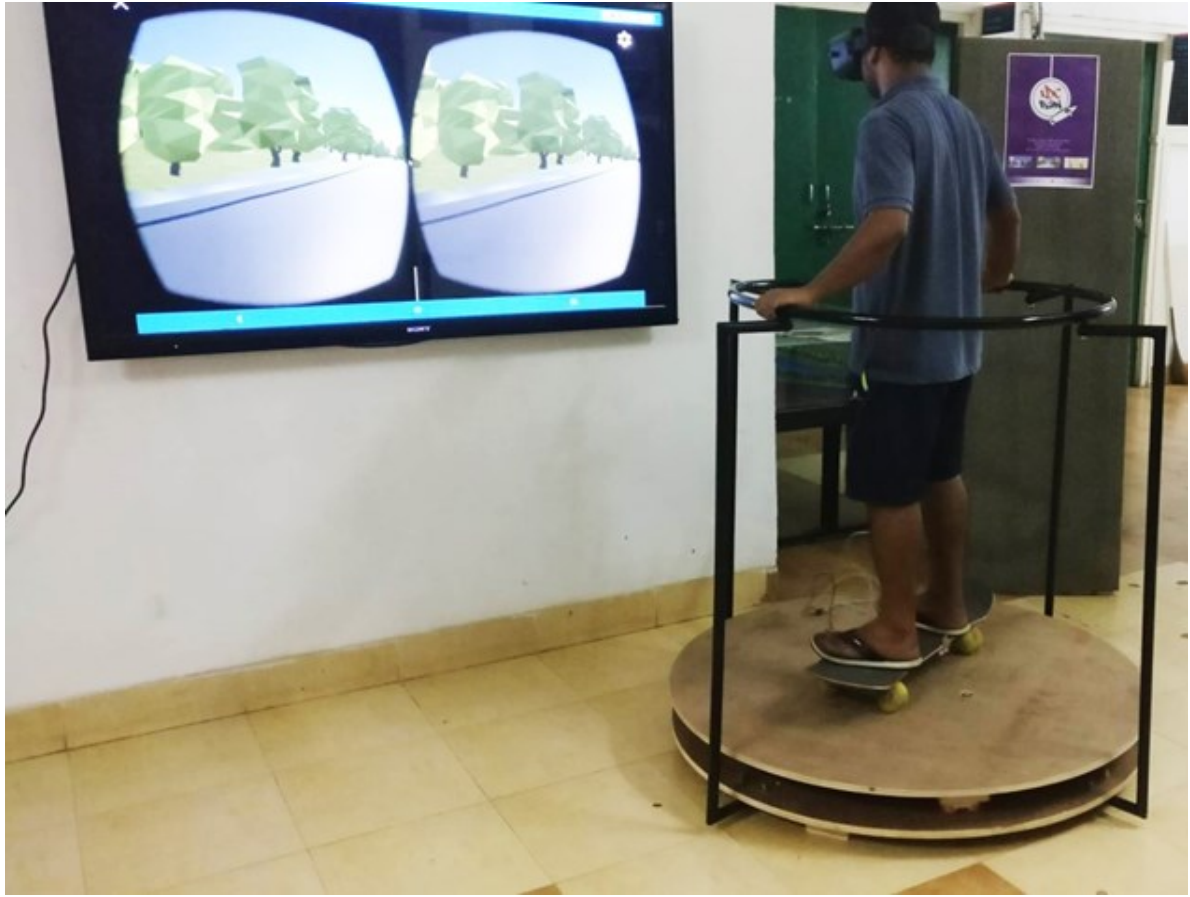

Source : Author

It allowed the player to turn and also maintain balance while performing complex gestures.

Players felt that the game was missing the real world laws of motion or physics. They didn't have a control of motion/ speed in which the skateboard should move.

A study of the movements was done to observe in detail how players move and how the skateboard decelerates. This concepts needed to be put into the controller design to mimic the haptic feedback of pushing the ground back in real world and the player moving forward standing on his skateboard in the virtual world.

An attempt to solve this problem was also made by Sato Daiki et al from Tokyo Institute of Technology by introducing a treadmill for one leg. The sensors to measure the movement and thrust of the push were embedded in the treadmill. Due to this, the readings were only taken depending on the physical rolling of the sensor rollers which were not very accurate. The electronic sensors, becoming a part of a bulky treadmill, defies the concept of portability and made the system expensive.

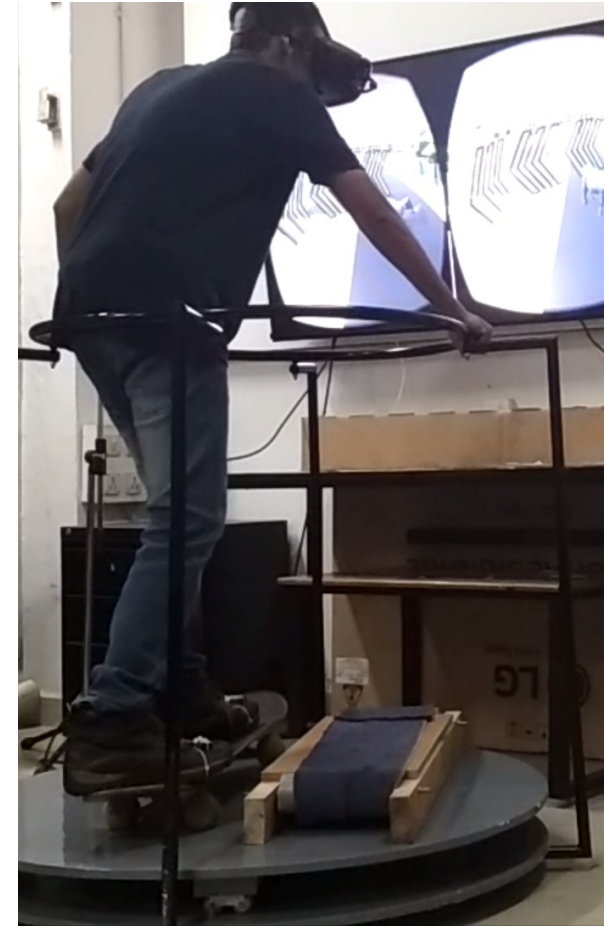

Source : Author

## 6.2. Final controller design prototype

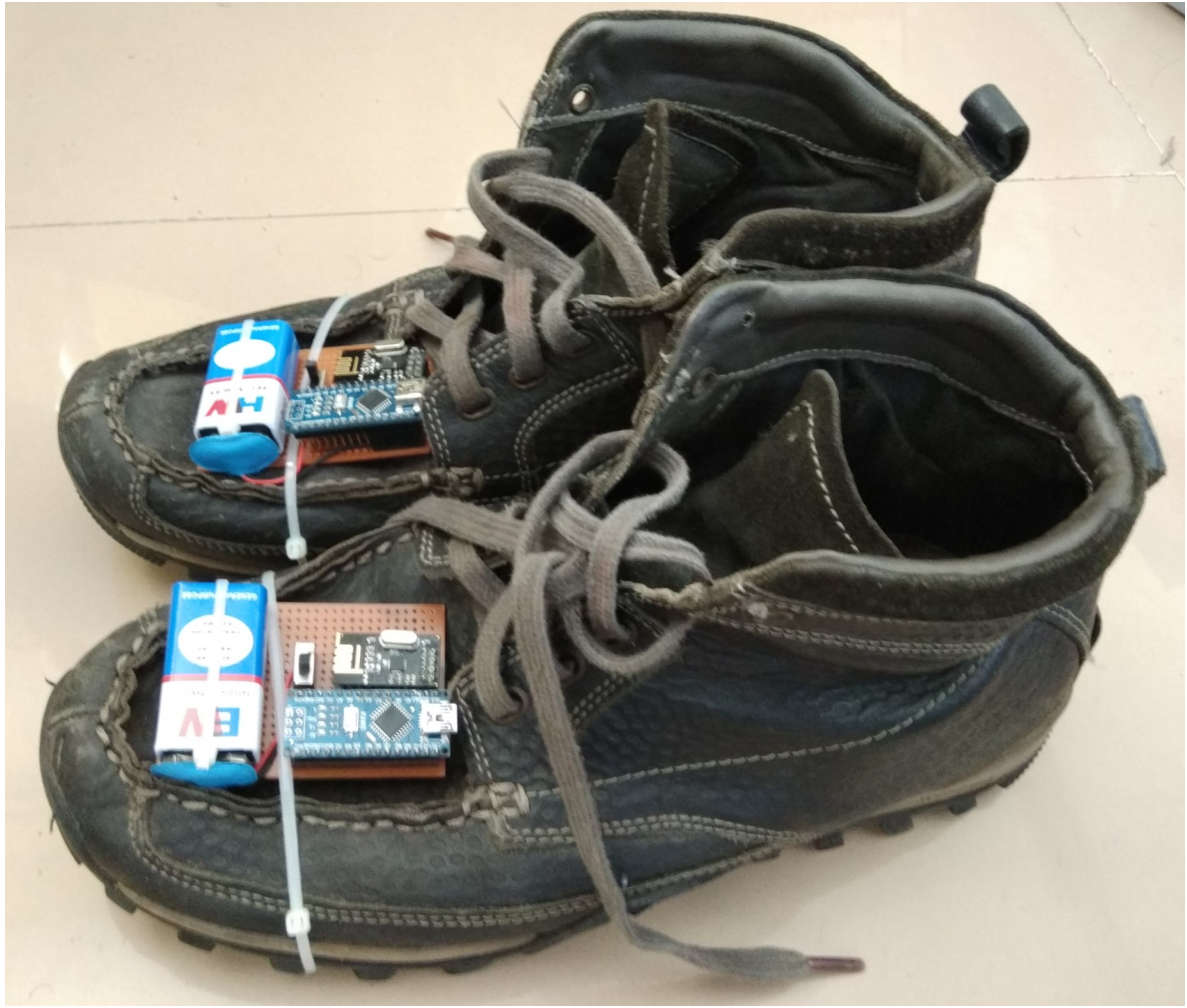

Source : Author

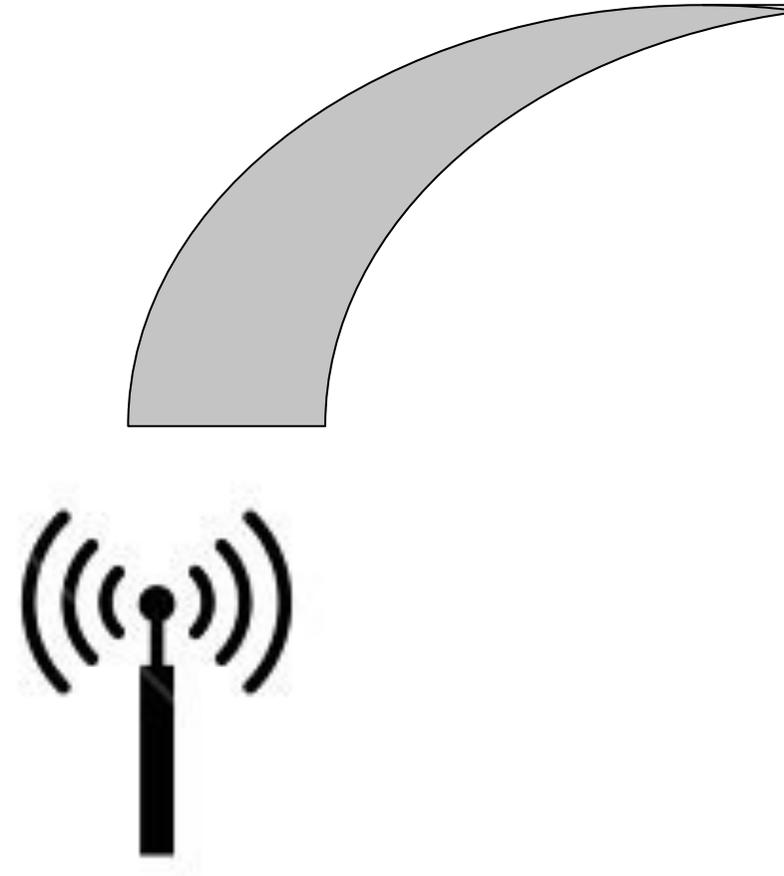

**Two transmitting units mounted on two shoes**

While trying to implement these in my controller, I felt it would be more easy to put these sensors as wearables on shoes than embedding sensors in skateboard and treadmill. The left leg should

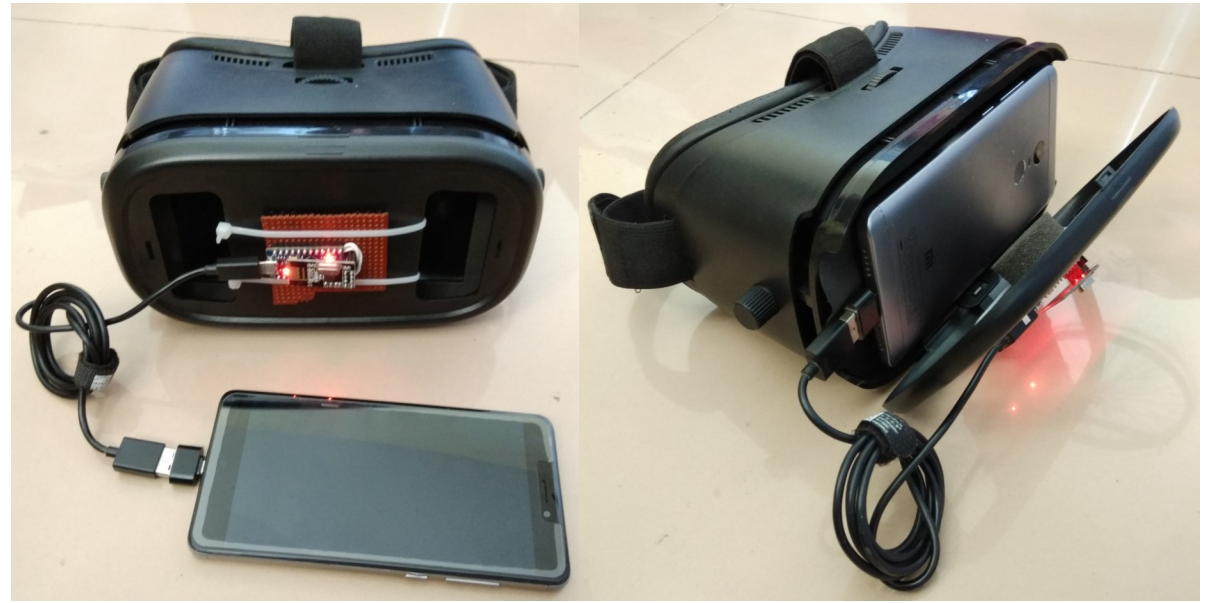

Source : Author

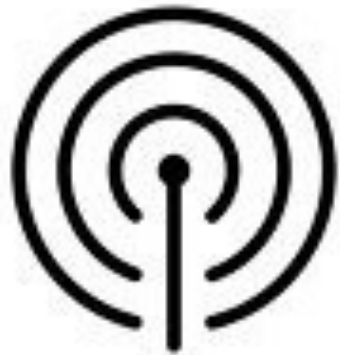

**Receiving unit mounted on HMD**

have the direction and jump sensing, while the right leg should have the ground push sensing. Two similar sensor modules were made with different thresholds of angle of gyration for sensing vertical, horizontal and lateral movements of two legs.

These data would be transmitted from the sensor modules to the receiver module, connected to the Smartphone or HMD. They will be processed by the game engine to render the virtual environment in real time. It eliminates the wiring of the HMD to the sensors,

The player can now be hands free with the controller. To engage the player, he can now control the turning/ rotation / pointing the direction of movement by turning the turntable platform underneath by applying angular force by hands.

In this way, the player has an increased sense of control/ interactivity with virtual world.

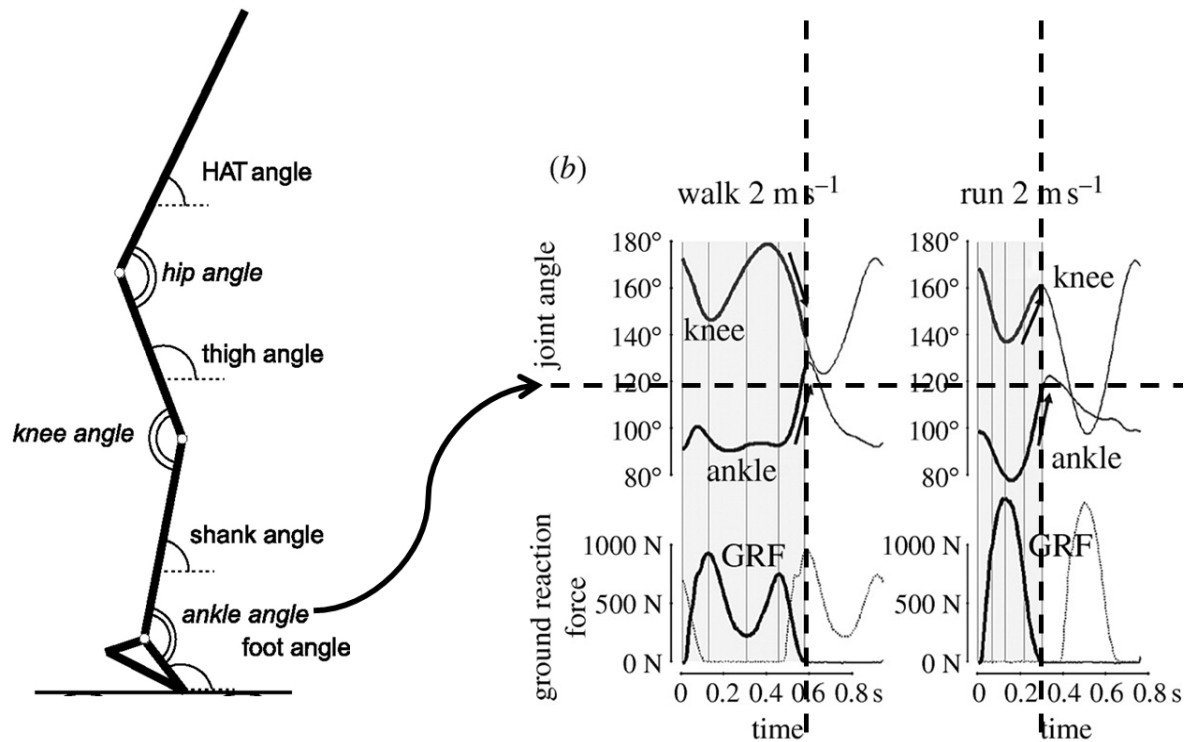

Source : Intelligence by machines

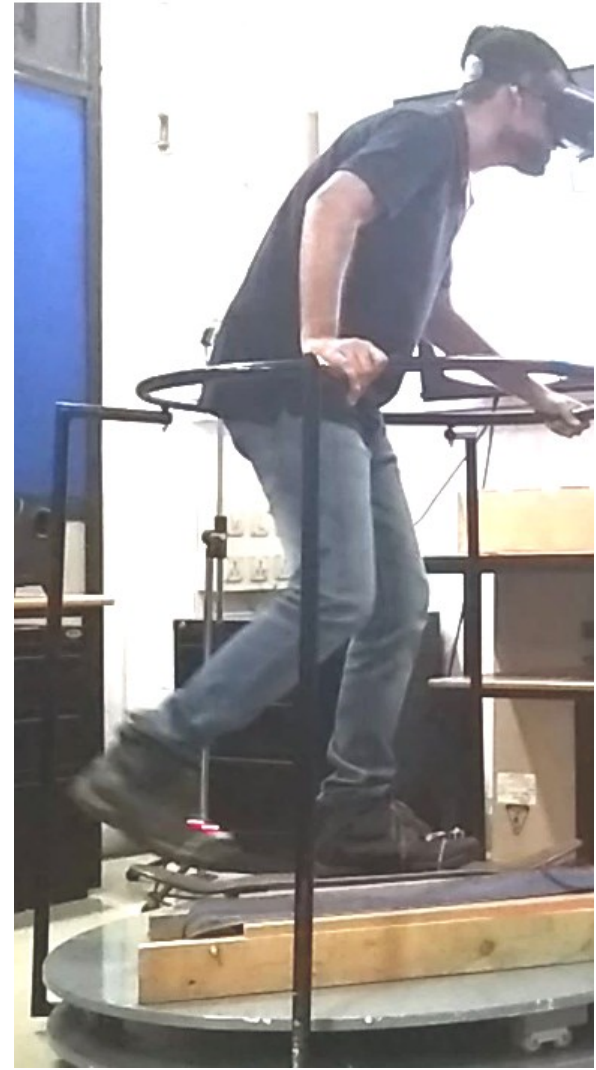

Source : Author

## 6.3. Ergonomic considerations

While pushing the leg back against the treadmill, there should be a position at which the sensor would detect the threshold of angle of gyration.

To determine the value existing data of angles of knee and ankle joint was plotted against time. We can see that at 120 degrees, the flow of action is

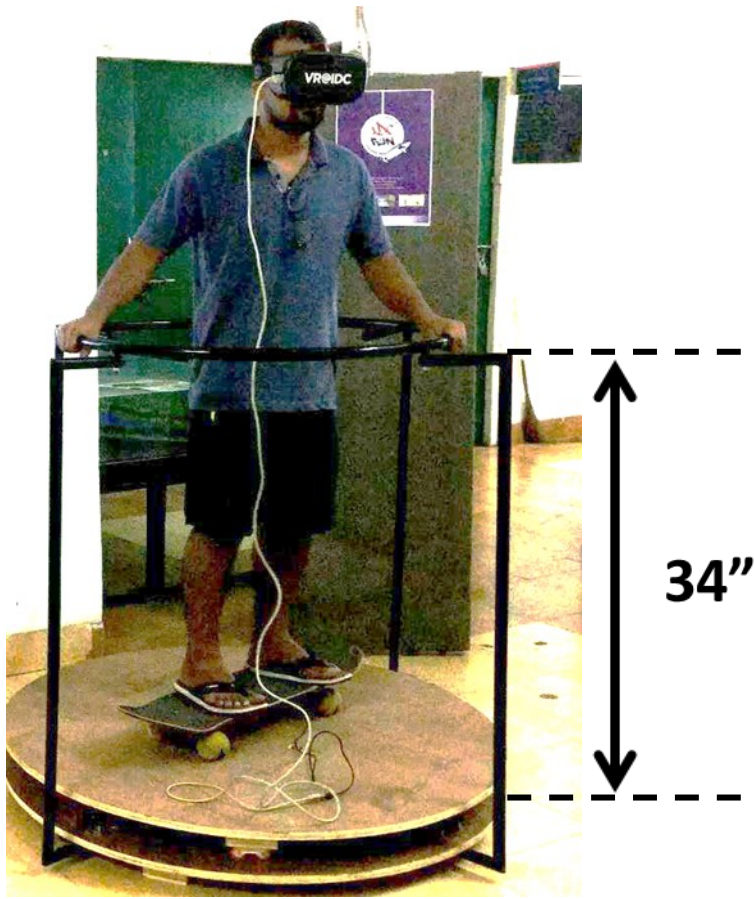

Source : Author

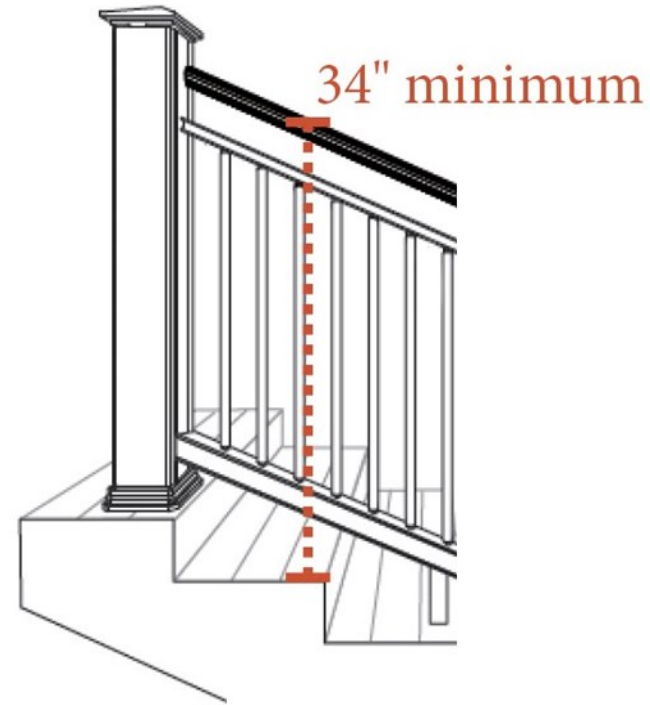

Source : Neufert Archhitect's Data 4th Edition

reversed. Corresponding to that, the force on the joints are also the least at this point. So we choose 120 degree as the threshold angle of running sensing.

As per Neufert's standards, the minimum height of hand rail support is 34 inches. It ensures that it is comfortable for tall people and also accessible to children.

# 7. Designing game environment

The potential of the IDC Run game is to operate as an effective and intuitive tangible VR interface, controlled by the movements of the player. The player here is a student from IIT Bombay. The objective of the game is to reach the department from hostel area as fast as possible by skating on a skateboard, while avoiding obstacles and collecting bonus coins.  The game environment was modelled keeping in mind the campus map of IIT Bombay and the actual architecture of buildings, roads, corridors, trees and other spaces.

The interactive movements in the game are designed to feel intuitive and enhance immersiveness. In this game the user's skateboard is mimicked in the virtual environment. The controller skateboard is embedded with ultrasonic sensors which controls the tilt and shift of the player. The player bends down or stands up to change the velocity of the skating movement. To jump, the player need to actually tilt the skateboard, the gesture actually used for jumping in real world. The player can jump over obstacles on the road and collect coins. Colliders at the roadside prevent player to cross the road borders and enter beyond that.

It is likely that the game has an ability to engage a player for more than a few minutes.

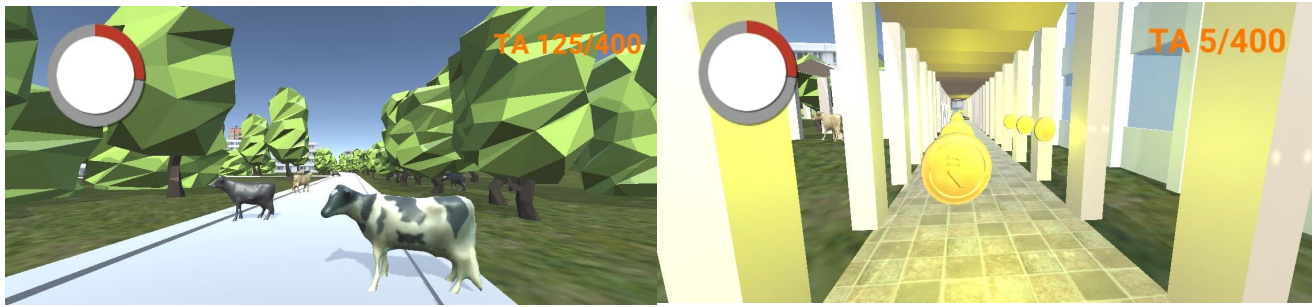

Source : Author

# 8. Usability evaluation

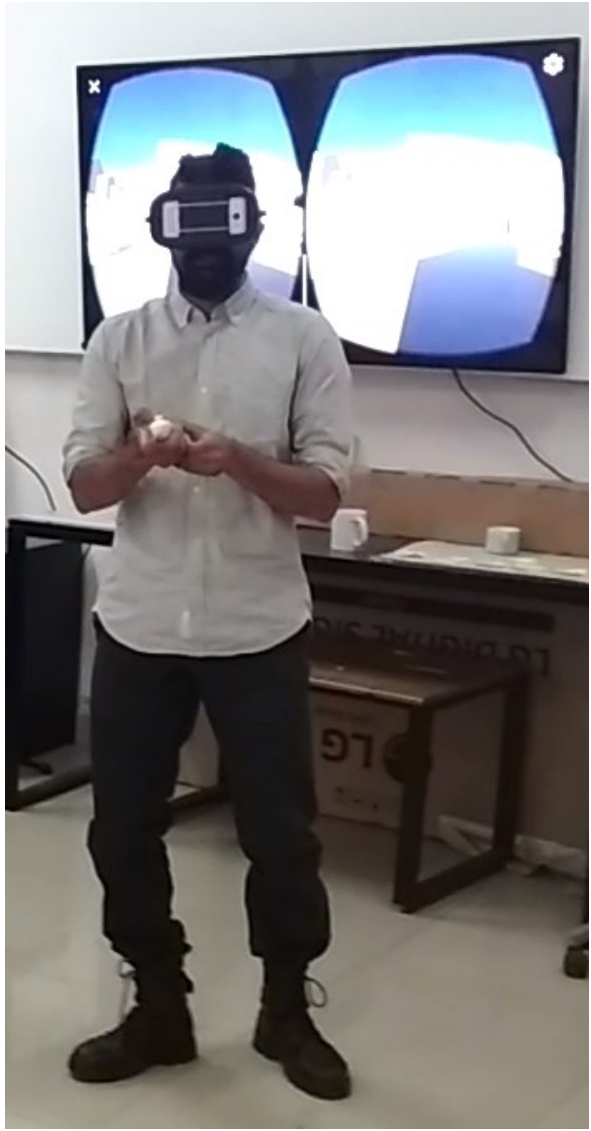

Source : Author

The wearable controller designed was tested for usability and compared with "Nun chuck Bluetooth controller" which is already available in market.

Usability study of the controller has been conducted formally with students after, them agreeing to sign the consent form (as per standards) and fill the response form ( Google Form )attached in this report.

## 8.1. Evaluation protocol

Participants all over IDC Department were called for Usability testing. 30 participants were tested for both the controllers on two different sessions (Total 60 tests), in consecutive days . Half of them were tested with one controller first and then the other and the vice-versa for the other half of the participants. It was expected that, it would reduce the experience and feedback bias of one experience over the other. Out of 30 participants, 18 were boys and 12 were girls. All participants were of age between 23 to 30 (mostly 24).

All the participants were given the task of going from Hostel15 to IDC and return back to the hostel in the VR game, IDC Skate. These sessions were recorded to deeply analyze, the activity and movement patterns with time, the type and frequencies of problems occurring due to various reasons (controller, communication, graphics, 3D environment, etc.).

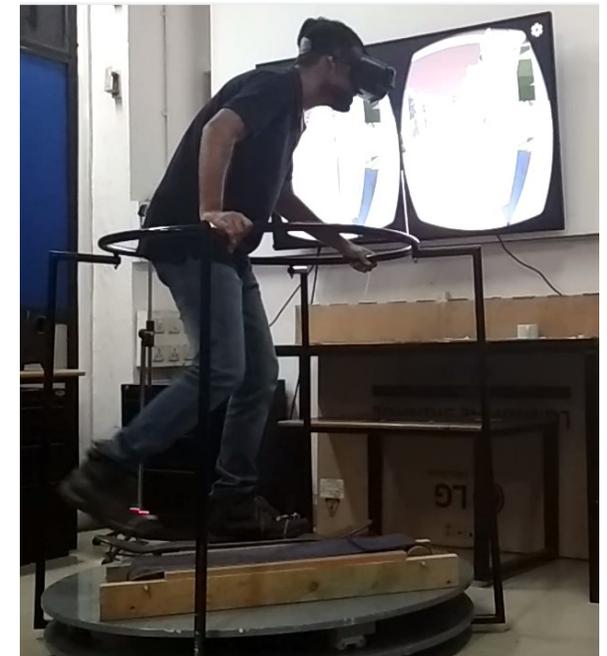

Source : Author

The answers to the questionnaire were analyzed and compared for hypothesis testing. In total there were 23 questions and the feedback was taken in Likert scale of 5 options from "Not Much" to "Too Much". The ordinal data was then tabulated in values 1-5 for an analysis of means. The means and SD for all participants in both the group of controllers were calculated. The difference between the means of the designed controller and the one, already available were compared. The sum of the differences depicts the overall effectiveness in experience of playing with the controller.

There is a marked difference between the means and the sum of overall differences, which proves the hypothesis :
The designed controller with natural gestures have more immersiveness in virtual reality.

|  | Nunchuck | | IDC Skate controller (Designed Controller) | | Difference (controller-nunchuck) |
| --- | --- | --- | --- | --- | --- |
|  | Mean | SD | Mean | SD |  |
| How much were you able to control events ? | 2.67 | 0.52 | 3.11 | 1.05 | 0.44 |
| How responsive was the environment to actions that you initiated (or performed)? | 2.33 | 0.82 | 3.16 | 1.21 | 0.82 |
| How natural did your interactions with the environment seem? | 2.68 | 1.16 | 3.50 | 0.63 | 0.82 |
| How natural was the mechanism which controlled movement through the environment? | 2.11 | 1.15 | 3.17 | 0.98 | 1.06 |
| Reaction to the real world disturbances ? | 2.79 | 1.13 | 2.83 | 0.41 | 0.04 |
| How aware were you of your control device? | 2.33 | 0.82 | 2.79 | 1.03 | 0.46 |
| How consistent or connected was the information coming from your various senses? | 3.47 | 1.12 | 2.50 | 1.05 | -0.97 |
| How much did your experiences in the virtual environment seem consistent with your real-world experiences? | 3.00 | 0.63 | 3.11 | 0.81 | 0.11 |
| Were you able to anticipate what would happen next in response to the actions that you performed? | 2.50 | 0.55 | 2.63 | 1.12 | 0.13 |
| How compelling was your sense of moving around inside the virtual environment? | 3.32 | 0.75 | 3.50 | 0.55 | 0.18 |
| How involved were you in the virtual environment experience? | 3.00 | 0.75 | 3.00 | 0.63 | 0.00 |
| Was the control mechanism non-distracting ? | 3.21 | 1.03 | 3.67 | 1.21 | 0.46 |
| Experience of delay between your actions and expected outcomes? | 2.79 | 1.13 | 2.00 | 0.63 | -0.79 |
| How quickly did you adjust to the virtual environment experience? | 2.84 | 1.12 | 3.17 | 0.75 | 0.32 |
| How proficient in moving and interacting with the virtual environment did you feel at the end of the experience? | 3.00 | 1.26 | 3.42 | 0.90 | 0.42 |
| Experience of interfere or distract you felt in visual display quality while performing assigned tasks or required activities? | 3.17 | 0.75 | 3.32 | 1.06 | 0.15 |
| Reaction due to the interfere of the control devices with the performance of assigned tasks or with other activities? | 2.58 | 1.02 | 2.83 | 0.75 | 0.25 |
| How well could you concentrate on the assigned tasks or required activities rather than on the mechanisms used to perform those tasks or activities? | 2.50 | 0.55 | 3.16 | 1.01 | 0.66 |
| Did you learn new techniques that enabled you to improve your performance? | 3.17 | 1.17 | 3.42 | 0.90 | 0.25 |
| Were you involved in the experimental task to the extent that you lost track of time? | 3.32 | 0.82 | 3.83 | 0.98 | 0.52 |
|  |  |  | Sum of Difference | | 5.33 |

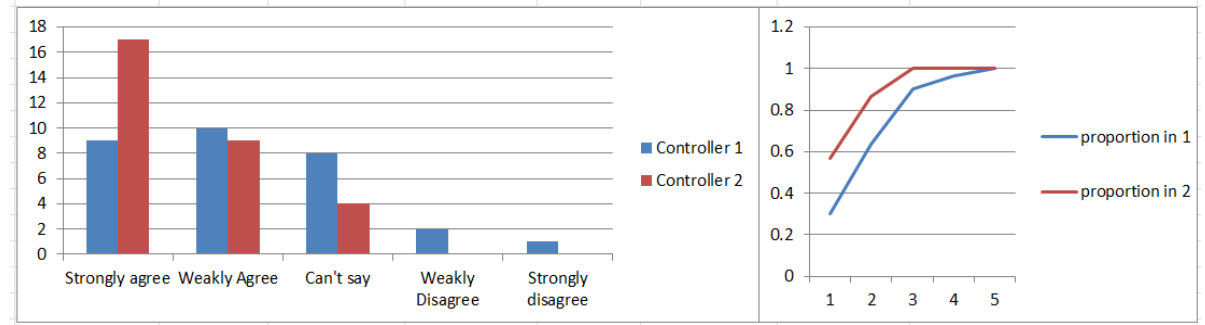

| | Observed in Controller 1 | Cumulative in 1 | Cumulative proportion in 1 | Observed in Controller 2 | Cumulative in 2 | Cumulative proportion in 2 | Absolute difference in proportions |
|---|---|---|---|---|---|---|---|
| Strongly agree | 9 | 9 | 0.3 | 17 | 17 | 0.566666667 | 0.266666667 |
| Weakly Agree | 10 | 19 | 0.633333333 | 9 | 26 | 0.866666667 | 0.233333333 |
| Can't say | 8 | 27 | 0.9 | 4 | 30 | 1 | 0.1 |
| Weakly Disagree | 2 | 29 | 0.966666667 | 0 | 30 | 1 | 0.033333333 |
| Strongly disagree | 1 | 30 | 1 | 0 | 30 | 1 | 0 |
| n | 30 | | | | | | 0.266666667 |
| Categories | 5 | | | | | | D Static (Highest of absolute diff.) |

**H0:** The results of testing of controllers (1) and (2) are significantly different from each other (Null hypothesis)

**H1:** The results of testing of controllers (1) and (2) are not significantly different from each other (Alternate hypothesis)

In this test, the Nun chuck controller (1) and the designed controller (2) were given to the same 30 users to play the game in two sessions. Half of them were tested with one controller first and then the other and the vice-versa for the other half of the participants. (All other conditions being the same). Then they marked the enjoyment in Likert scale of 5 values.

The D value is calculated from Kolmogorov- Smirnov test and checked against the D value table.

While level of significance is 0.01 (99% confidence), D static exceeds the table value thus hypothesis gets rejected. While level of significance is 0.05 (95% confidence) or is 0.1(90% confidence) D static exceeds the table value thus hypothesis gets rejected.

From the last two experiments, we can say that the designed controller (2) is significantly better than (1), the Nunchuck controller available in market with a confidence of 95%.

# 9. User Feedback

Though, the controller is under significant refinement process, but the assessment is done as per its completion till date.

It was first presented in a public demo booth in HCI conference, Interact 2017. The controller was exhibited, near a wide range of other VR game demos. Exceptionally compelling was the level of engagement that players had with the amusement, with numerous players not having any desire to quit playing. It was a good response, though the functionality of the game till now is very limited. It was intriguing to observe the players being energetic and cheerful.

The controller rather than being considered as a communicator or an external entity, it was successful in initiating movement and interaction very naturally. In general, players were able to figure out how to use the controller in the game without asking for instructions. It leads towards the fact that the interface and controls are reasonably intuitive.

It was observed that, the control of the direction to move was figured out more easily than gestures like jumping. Interviewed responses from the players was compelling. Players said that the physical controls were effective in intriguing interaction.

Feedback from players confirmed that the controller had gone a long route towards accomplishing its objective.

Considering, the controller was tested with students from IDC and not with people with diverse backgrounds, we cannot reliably judge the product until a formal rigorous study is done with a larger population.

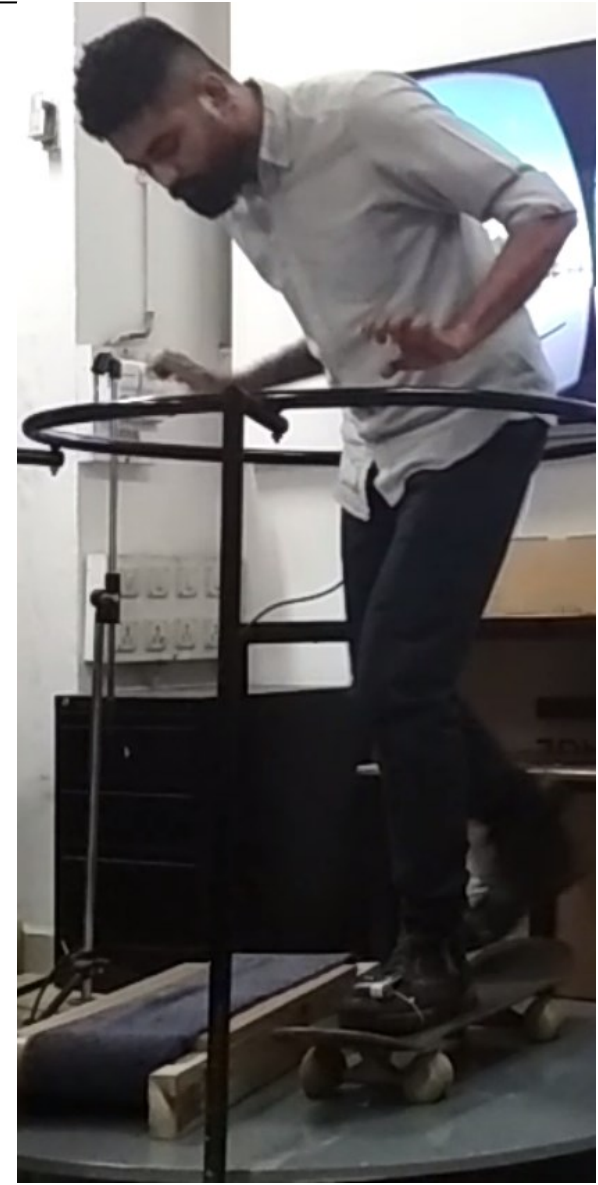

Source : Author

# 10. Takeaway for future VR Controller designers

This project has a lot more to contribute than an open source code for programming sensors and controllers.

- The project presents a story of how to conduct surveys from users, bring out critical data, analyze them and decide how to use them in designing the product/ part of the overall ecosystem. Every kind of data obtained is a bit different. The process of deciding critical dimensions required for designing specific gestures from a sample data following ergonomics guidelines, is something the future design projects can refer to.

- Working on this kind of project requires a keen interest in multiple domains like design, electronics, coding etc. The process of deciding the right electronic components *(microprocessor and sensors)* for sensing specific human gestures leads to a better design decision in designing the natural interactive gesture, the form factor of the product and its usability.

- Different ways of putting the sensors/controllers were tried out throughout the project. Some of it were embedded and some were wearables. The process of deciding the appropriate way to mount electronic sensors / devices to sense useful data, would yield good data required to perform specific task. It can also be considered in broad spectrum of designing where and how the product fits and enhances usability.

- It is very hard to compare and contrast, competitive products in the market. This project shows how one can go through a scientific way to evaluate and compare the product's usability and immersiveness through the product. Collecting very subjective choices/ answers from users as feedback and putting it in an evaluation format for a critical analysis of measuring differences between them, should be reused as a model by other product/ interaction designers/developers.

Apart from these most important/general ones, there are more to learn from this project irrespective of the background one belongs to. I believe this project would help game designers in initiating new thoughts.

# 11. Concluding thoughts

The physical skateboard controller project has explored an alternative method of interaction with HMDs by creating an artefact capable of providing physical gesture based input and getting real time immersive experience. The potential applications of such technology is widespread and not limited to skateboard game controllers.

The emphasis of the project was to create a device that seamlessly bridges the user experience "in the real world" to the actions in the virtual world. It challenges the notion of what a tangible interface is, by becoming an invisible mediator between the human and the computer/HMD, yet having physical engagement potential. This enables intuitiveness for the user. As this technology develops further, the interactions between real and virtual players will very closely resemble each other. This has the potential to improve the immersion of virtual experiences. There is even the potential for this technology to allow for new methods of experiences that are yet to be discovered.

Virtual reality is currently going through a phase of worldwide excitement, and both industry and academic researchers curiously exploring ways in which sports can be played within this new kind of lean in VR environment.

# USER STUDIES : IDC SKATE

## Consent Form

I agree to participate in the usability study conducted by Arka Majhi for the game and controllers designed by him: **IDC Skate**

I understand that participation in this usability study is voluntary and I agree to immediately raise any concerns or areas of discomfort during the session with the study administrator.

Please sign below to indicate that you have read and you understand the information on this form and that any questions you might have about the session have been answered.

**Date:** _________

**Please write your name:** ______________________________________________

**Please sign your name:** ______________________________________________

**Thank you!**

We appreciate your participation.

# User Testing : Controllers

Thank you for participating in the User Testing session. I want to hear your feedback so that I can keep improving my product. Please fill this quick survey and let me know your thoughts.

* Required

**Email address ***

Your email

**Name ***

Your answer

**Sex ***
- ○ Female
- ○ Male
- ○ Other
- ○ Prefer not to say

**Age ***

Choose ▼

**Course ***

Choose ▼

**Contact no. ***
(Required to call you for the next test. Or if some response found invalid/skipped. No spam guaranteed)

Your answer

**Controller you played with ***

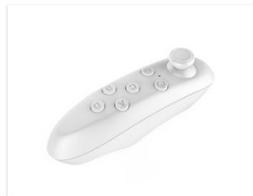
○ Nunchuck

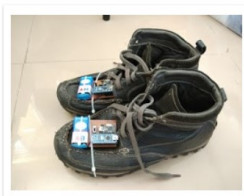
○ IDC Skate Platform

**It was very interactive ***
(Each and every actions/gestures of mine were mimicked flawlessly in the virtual world)

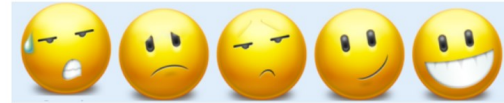

Strongly disagree — 1 2 3 4 5 — Strongly agree

**It was very real ***
(The virtual world felt very much like the real world)

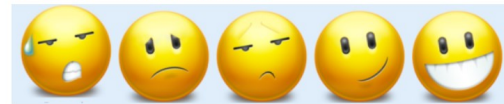

Strongly disagree — 1 2 3 4 5 — Strongly agree

**I enjoyed a lot ***
(Overall, the experence was very enjoyable)

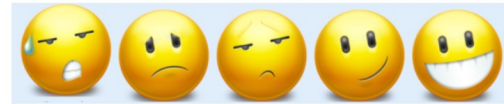

Strongly disagree — 1 2 3 4 5 — Strongly agree

**How much were you able to control events ? ***

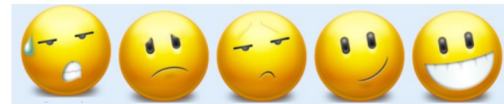

Not much — 1 2 3 4 5 — Too much

**How responsive was the environment to actions that you initiated (or performed)? ***

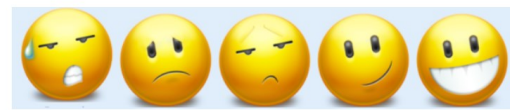

Not much — 1 2 3 4 5 — Too much

**How natural did your interactions with the environment seem? ***

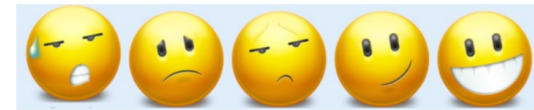

Not much — 1 2 3 4 5 — Too much

**How completely were all of your senses engaged? ***

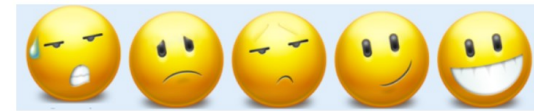

Not much — 1 2 3 4 5 — Too much

**How natural was the mechanism which controlled movement through the environment? ***

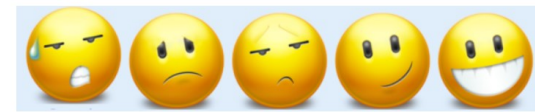

Not much — 1 2 3 4 5 — Too much

**Reaction to the real world disturbances ? ***

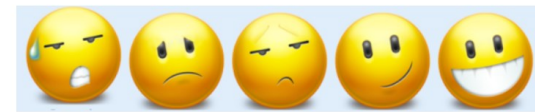

Not much — 1 2 3 4 5 — Too much

**How aware were you of your control device? ***

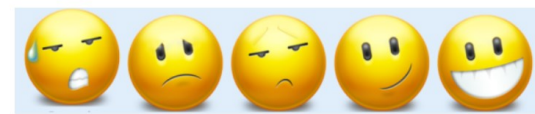

Not much — 1 2 3 4 5 — Too much

How consistent or connected was the information coming from your various senses? *

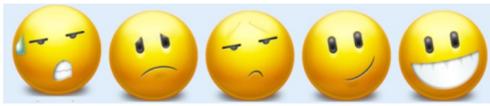

Not much  1  2  3  4  5  Too much

How much did your experiences in the virtual environment seem consistent with your real-world experiences? *

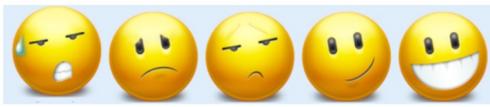

Not much  1  2  3  4  5  Too much

Were you able to anticipate what would happen next in response to the actions that you performed? *

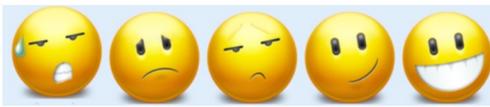

Not much  1  2  3  4  5  Too much

How compelling was your sense of moving around inside the virtual environment? *

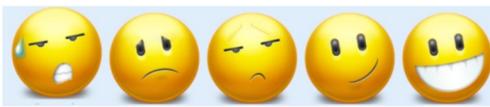

Not much  1  2  3  4  5  Too much

How involved were you in the virtual environment experience? *

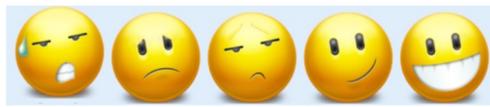

Not much  1  2  3  4  5  Too much

Was the control mechanism non-distracting ? *

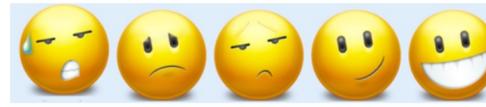

Not much  1  2  3  4  5  Too much

Experience of delay between your actions and expected outcomes? *

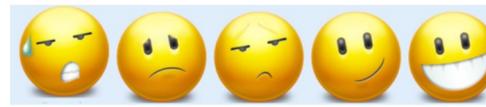

Not much  1  2  3  4  5  Too much

How quickly did you adjust to the virtual environment experience? *

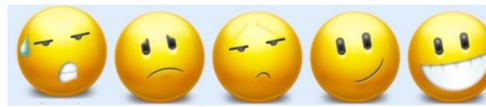

Not much  1  2  3  4  5  Too much

How proficient in moving and interacting with the virtual environment did you feel at the end of the experience? *

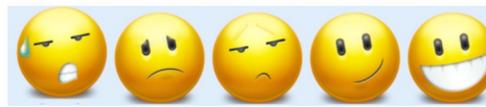

Not much  1  2  3  4  5  Too much

Experience of interfere or distract you felt in visual display quality while performing assigned tasks or required activities? *

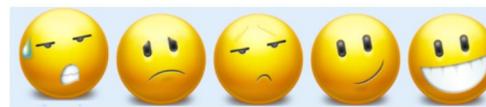

Not much  1  2  3  4  5  Too much

Reaction due to the interfere of the control devices with the performance of assigned tasks or with other activities? *

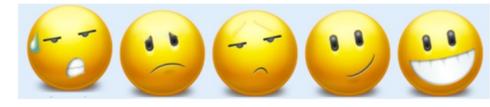

Not much  1  2  3  4  5  Too much

How well could you concentrate on the assigned tasks or required activities rather than on the mechanisms used to perform those tasks or activities? *

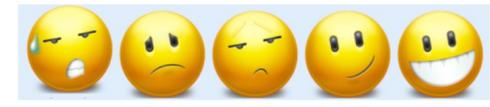

Not much  1  2  3  4  5  Too much

Did you learn new techniques that enabled you to improve your performance? *

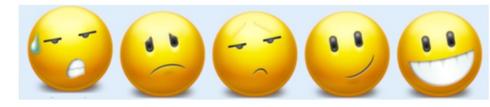

Not much  1  2  3  4  5  Too much

Were you involved in the experimental task to the extent that you lost track of time? *

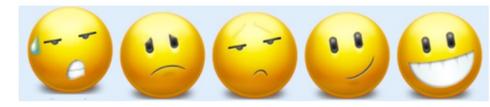

Not much  1  2  3  4  5  Too much

Which aspect/(s) overall did you enjoy the most ? *

Your answer

Which aspect/(s) of the system do you think need improvement ? *

Your answer

Which aspect/(s) you hated the most ? *

Your answer

33